\documentclass[twocolumn,floatfix,superscriptaddress]{revtex4}
\usepackage{color}

\usepackage{graphicx}
\usepackage{amsmath}
\usepackage{latexsym}
\usepackage{epstopdf}
\usepackage{amsmath}
\usepackage{amssymb,amsmath}
\usepackage{multirow}
\usepackage{mathtools}
\newcommand{\beq}{\begin{equation}}
\newcommand{\eeq}{\end{equation}}

\begin{document}

\title{Universality and conformal non-invariance in self-affine rough surfaces}
\author{S. Hosseinabadi,$^1$ M.~A.~Rajabpour,$^2$ M. Sadegh Movahed,$^{3,4,6}$, S. M. Vaez Allaei$^{5,6}$\\
$^1$ Department of Physics, East Tehran Branch, Islamic Azad University, Tehran, Iran\\
$^2$ Instituto de F\'{\i}sica de S\~{a}o Carlos, Universidade de S\~{a}o Paulo, Caixa Postal 369, 13560-590, S\~{a}o Carlos, SP, Brazil\\
$^3$ Department of Physics, Shahid Beheshti University, G.C., Evin, Tehran 19839, Iran\\
$^4$ School of Astronomy, Institute for Research in Fundamental Sciences, (IPM), P. O. Box 19395-5531, Tehran, Iran\\
$^5$ Department of Physics, University of Tehran, Tehran 14395-547, Iran\\
$^6$ The Abdus Salam International Centre for Theoretical Physics, Strada Costiera 11, I-34013 Trieste, Italy\\}

\vskip 1cm

\begin{abstract}
 We show numerically that the  roughness  and growth
 exponents of a wide range of rough surfaces, such as random
deposition with relaxation (RDR), ballistic deposition (BD) and
restricted solid-on-solid model (RSOS),   are independent of the
underlying regular (square, triangular, honeycomb) or random
(Voronoi) lattices. In addition we show that the universality holds
also at the level of statistical properties of the iso-height lines
on different lattices. This universality is revealed by calculating
the fractal dimension, loop correlation exponent and the length
distribution exponent of the individual contours. We also indicate
that the hyperscaling relations are valid for the iso-height lines
of all the studied Gaussian and non-Gaussian self-affine rough
surfaces. Finally using the direct method of Langlands et.al we
show that  the contour lines of
the rough surfaces are not conformally invariant except when we have
simple Gaussian free field theory with zero roughness exponent.\end{abstract}
\maketitle
\section{Introduction}

Many theoretical and numerical efforts have been focused on the study, characterization,
and understanding of stochastic surface patterns, for various growth models relevant to
 non-equilibrium processes \cite{Vicsek,Barabasi,Meakin}.
The surface roughening phenomena have been intensively studied via various discrete models
and continuum equations. Scaling properties have been observed in time and space fluctuations
of these surfaces and such interfaces show self-similar or self-affine properties \cite{Meakin}.

Various growth models are often characterized and classified by three exponents, the roughness
 exponent, $\alpha$, the dynamical exponent, $z$, and the growth exponent, $\beta$. Most of the
work is thus devoted to identifying the different universality classes to which the models studied
belong \cite{EW,Family,KPZ,Meakin1,Halpin,BD1,BD2,RSOS}. For example various discrete growth models
such as  the ballistic deposition (BD) \cite{BD1,BD2}, Eden \cite{Family} and the restricted solid-on-solid
(RSOS) \cite{RSOS} models were known to be described by the Kardar-Parisi-Zhang (KPZ) equation in one
dimension \cite{KPZ}. Random deposition with surface relaxation (RDR) is another important discrete growth model
 that can be described by the Edwards-Wilkinson (EW) equation \cite{EW}.

One of the most popular methods for  characterization of the rough
surfaces is based on the concept of fractal properties of iso-height
lines called contour lines. The contour plot consists of closed
non-intersecting loops in the plane that connect points of equal
heights. Several experimental and numerical studies obtained
 the characterization of the fractal properties of loop ensembles of (2+1)Dimensional self-affine rough
 surfaces such as glassy interfaces and turbulence \cite{kondev1}, (2+1)Dimensional fractional Brownian
 motion \cite{smvaez}, discrete scale invariant rough surfaces \cite{ghasemi}, KPZ surfaces \cite{Niry},
 the multi-fractal surfaces \cite{multi},  experimental data coming from the AFM analysis of WO$_3$
 surfaces \cite{rajabpour} and also STM images of rough metal surfaces \cite{kondev2}. One of the most
 important finding in this direction is the dependence of
the different exponents i.e.  the fractal dimension of one contour,
$D_f$, the length distribution exponent, $\tau$, and the loop
correlation-function exponent, $x_l$, related to the contour lines of
mono-fractal rough surfaces to the only universal parameter in the
system, i.e. roughness (Hurst) exponent. This conjecture confirmed
by many numerical simulations but there is no theoretical proof yet
\cite{kondev2}.

Most of the  discrete surface growth models in (2+1)Dimension have been
simulated on the square lattice with periodic boundary conditions
with the topology of the torus \cite{Barabasi}. There are a few
works devoted to the study of the dynamical scaling exponents of
surface growth  on substrates with fractal structures
\cite{fracsub1,fracsub2}. Here we  numerically study  the discrete
growth models including  random deposition (RD), RDR, BD and RSOS on
different two dimensional lattice types such as square ($\mathcal{S}$), honeycomb  ($\mathcal{H}$),
triangular )$\mathcal{T}$), and also Voronoi random structure ($\mathcal{V}$). We show that
universality holds, independent of the lattice type.  The dynamical
exponents $\alpha$ and $\beta$ for the surface growth process and
also the geometrical scaling exponents
 $D_f$, $x_l$ and $\tau$ for the contour lines of such surfaces are not affected by the change of
 the substrate's lattice type.

Another method to study two dimensional critical systems is by
Schramm-Loewner evolution (SLE$_\kappa$). It gives a powerful tool
to classify all the conformally invariant curves in two dimensions,
for review
 see \cite{Baur}. Recently, some studies argued that scaling exponents of 2D systems i.e. zero-vorticity
 lines of Navier-Stokes turbulence \cite{turbulence} and domain walls in statistical models \cite{Baur}
 can be determined by the diffusivity constant $\kappa$. We will show that the contour loop ensemble of RDR model,
  independent of lattice structure is conformal invariant object and could be described by SLE$_4$. Our numerical
   calculations show that  the contour lines of RSOS model are not conformal invariant objects,  however they can still possibly be classified by
   the Loewner's equation.

The structure of this paper is organized as follows: In the next
section we will investigate the scaling relations for the discrete
surface growth processes. Scaling behaviors and corresponding
properties of contour lines are also presented in detail in this
section. The numerical results to measure scaling exponents for the
stochastic surfaces  i.e. RD, RDR, BD and RSOS models on different
lattice types are discussed in the third section. In this section
  we also identify geometrical properties of loop ensembles by means of conformal invariant test.
In the last section we will summarize our studies and we draw some concluding remarks.

\section{Scaling of rough surfaces}
There are many non-equilibrium surface growth processes which
exhibit scaling properties. Different models with the same scaling
exponents are grouped into universality classes characterized by the
same value of the critical exponents. Two methods to get some
information about the scaling properties of the surface growth
processes are $i)$ the dynamic evolution of the aggregate interface
and, $ii)$ iso-height lines of the surfaces in the saturation
regime.
\subsection{Dynamical scaling exponents}
The simplest quantitative behavior  of a given aggregate interface
is its interface width $W(t,L)\equiv \sqrt{\big\langle
\frac{1}{L^d}\sum_{i=1}^{L^d}(h_i(t)-\bar{h}(t))^2 \big\rangle }$
where $\bar{h}(t)$ is spatial average of height at time $t$ and $h_i(t)$ is the local height variable at the site $i$.  The
averaged roughness over different configurations  show  scaling
behaviors. The width is saturated as $W \propto t^{\beta}$ with
growth exponent $\beta$ and $W \propto L^{\alpha}$ with roughness
$\alpha$ for short time and long time limits, respectively. The
scaling exponents $\alpha$ and $\beta$ are used to characterize a
given universality class of surface growth process \cite{Barabasi}.
Another quantity that describes dynamic of the growing rough
surfaces is the height-height correlation function $C(r,t) \equiv \langle
\left[ h(\mathbf{r}_0+\mathbf{r},t)-h(\mathbf{r}_0,t) \right]^2
\rangle$. The heights separated by the short distances $r \ll L$ are
fully correlated and the correlation function scales as
$C(r,t) \propto r^{2\alpha}$ \cite{Barabasi}.

\subsection{Geometrical scaling exponents}
For a given scale invariant height configuration
$h(\mathbf{r})$, at the level cut at the mean height
$h(\mathbf{r})=h_0$, there are many closed non-intersecting
contour loops that connect points of equal height. A few scaling
functions and scaling exponents need to characterize the size
distribution of such contour lines. Contour loops with length $s$
and radius $R$ are scale invariant and follows a power law $\langle
s \rangle \sim \langle R\rangle^{D_f}$, where $D_f$ is the fractal
dimension of a contour loop. The contour line properties can be described
by the probability distribution of contour lengths $\tilde {P}(s)$.
This function measures the probability that one loop has length $s$
and it follows the scaling behavior $\tilde{P}(s)\sim s^{-\tau}$
where $\tau$ is a scaling exponent. Another interesting quantity
with the scaling property is the loop correlation function
$G_c(\mathbf{r})$. This function is a probability measure for the
two points in the plane where separated by the distance $\vert
\mathbf{r}\vert$  lie on the same contour. For iso-height lines on
the grid with lattice constant $\mathbf{c}$ and in the limit $\vert
\mathbf{r}\vert \gg \mathbf{c}$, the two point correlation function
is also scale invariant $G_c(\mathbf{r})\sim \frac{1}{\vert
\mathbf{r} \vert ^{2x_l}}$, where $x_l$ is the loop
correlation-function exponent. For the contour loops on a Gaussian
surface, $x_l=\frac{1}{2}$ is conjectured \cite{kondev3}. There is
no any mathematical proof for this hypothesis  yet. The estimated
value $x_l=\frac{1}{2}$ is confirmed by various numerical works for
large class of the known self-affine rough surfaces
\cite{kondev1,ghasemi,kondev2}.

For the Gaussian self-affine rough surfaces with the roughness
exponent $\alpha$ and the scaling exponents $D_f$, $\tau$ and $x_l$,
it was shown that the two well known hyper-scaling relations
\begin{equation} \label{hyperscaling}
D_f(\tau-1) = 2-\alpha, \hspace{1cm} D_f(\tau-3)=2x_l-2,
\end{equation}
are valid \cite{kondev2}.  Finally there is another measure associated with a given contour loop
ensemble which is called the two
point correlation function for contour lines with length
$s$, $G_s$. Scaling properties of the contours force $G_s(r)$ to scale
with $s$ and $r$ as $G_s(\mathbf{r})\sim s^m |\mathbf{r}|^{-n}f_{G_s}(\frac{r}{R})$ where the new exponents satisfy \cite{kondev1}
\begin{equation} \label{hyperscaling2}
2-n=D_f(1-m).
\end{equation}

In the next section we numerically calculate
all introduced exponents $\alpha$, $\beta$, $D_f$, $\tau$, $x_l$, $n$ and $m$
 for  RD, RDR, BD and RSOS on different lattice structures and we
will show the validity of the two scaling relations in Eq.
(\ref{hyperscaling}) and also the Eq. (\ref{hyperscaling2}).

\section{Numerical results}
The lattice version of RD, RDR, BD and RSOS models are simple to
describe. In all of them particles fall  on vertical direct line
onto the substrate from a random position above the surface
\cite{Barabasi}. In the simple on-top site RD model, the particle
deposits on randomly chosen site position. In RDR model a particle
is released on the top of a selected column, but it does not stick
to the surface immediately and it can diffuse to a nearest neighbor
site of lower height \cite{Barabasi,EW}. Surface diffusion generates
correlation between neighbor sites and it causes surface width
saturation effect. In BD model, falling particle sticks to the old
one either on the top or a nearest neighbor occupied site
\cite{Barabasi,BD1,BD2}. In RSOS surface growth process the particle
aggregates to the substrate if and only if the difference between
heights of all pairs of nearest-neighbor columns  satisfies the
condition $\Delta h \leq 1$; otherwise,  if this condition is not
met, the corresponding aggregation attempt is rejected from the
system \cite{RSOS}.

In order to find dynamical and geometrical scaling properties of RD,
RDR, BD and RSOS models on the initially flat substrates, we have
generated these models on different lattice types, i.e. honeycomb,
triangular, square and Voronoi with size $L=512$ and all
measurements are made using an ensemble of 2000 realizations (for BD
model the simulation were also done on lattices with size $L=2048$
with 500 samples). In order to reduce the errors due to the
substrate's boundaries, we used periodic boundary condition during
particle deposition in horizontal and vertical directions.
 Each time step is defined as the number of particles to fall up surface on the average,
 which is equal to $L\times L$. To extract the contour lines of the saturated surfaces at the mean height, $h_0$ we
 used Hoshen-Kopelman algorithm \cite{HK}.

\subsection{Distribution of heights}

To check that the surface that we have is a Gaussian surface or not we used the definition of local curvature at $\mathbf{r}$ and at scale $b$  which is \cite{kondev3}
\begin{equation} \label{local curvature}
C_b(\mathbf{r})\equiv \sum_{m=1}^M \left [h(\mathbf{r}+b \bold{e}_m)-h(\mathbf{r})\right ]
\end{equation}
where the sum of $\bold{e}_m$'s are a fixed set of vectors summing to zero. The distribution of the local curvature in Gaussian surfaces is Gaussian. In Fig. ~1 one can see 
that the distribution is Gaussian for RDR but non-Gaussian for RDR and BD. To quantify this claim we depicted in Fig. ~2
the third $\frac{\langle C_b(\mathbf{r})^3\rangle}{\langle C_b(\mathbf{r})^2\rangle^{3/2}}$ and fourth moments $\frac{\langle C_b(\mathbf{r})^4\rangle}{\langle C_b(\mathbf{r})^2\rangle^{2}}$ of the local curvature which should
be $0$ and  $3$ for a Gaussian surfaces respectively. Based on these two figures we conclude that among the surfaces that we are going to study just 
the RDR is a Gaussian surface. It is worth mentioning that by increasing the scale $b$, the results of RDR and BD 
surface growth models converge to the Gaussian process. We also repeated these
 computations for the other lattice types and the results are the same as that of for square 
lattice type. Finally we also checked the self-affinity of our surfaces by showing 
that for all the surfaces we have $\langle[C_b(\mathbf{r})]^q\rangle\sim b^{q\alpha}$.

 \begin{figure}
\begin{center}
\includegraphics[width=0.8\linewidth]{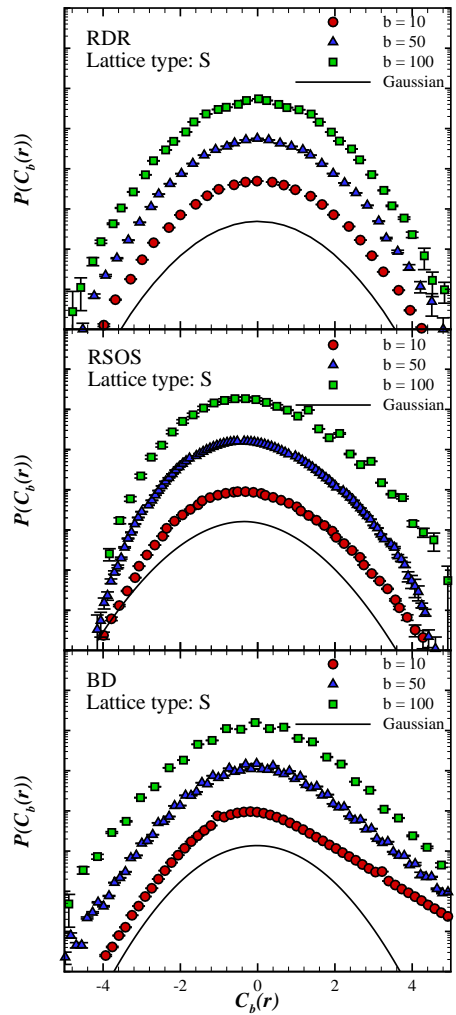}\\
\caption{\label{fig.1} (Color online)
 The distribution of the local curvature in RDR, BD and RSOS model. To make more sense we shifted various plots vertically. }
\end{center}
\end{figure}
 
\begin{figure}
\begin{center}
\includegraphics[width=0.8\linewidth]{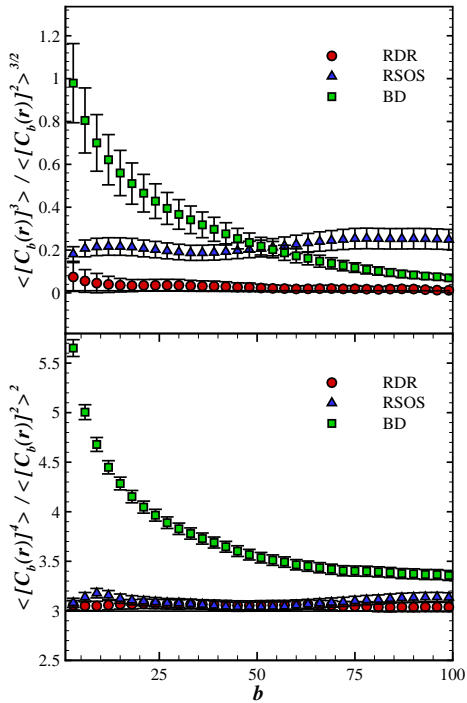}\\
\caption{\label{fig.2} (Color online)
 The third and fourth moments of the local curvature in RDR, BD and RSOS models on square lattice type.}
\end{center}
\end{figure}

\subsection{Scaling exponents}

One approach to show that two rough surface models belong to the same universality class is to compute all scaling exponents of the models.
In this sub-section we report measured values of the scaling exponent for various models.

\begin{itemize}
\item[a. ]\textit{The growth scaling exponent}
\end{itemize}

 In Table. \ref{tab1}, we have shown the results for
various growth models on different lattice types. Our measurements
of the scaling exponent $\beta$  on the square lattice are
consistent with previous studies \cite{Barabasi,saberi}.

\begin{table}[htp]
\caption{ Numerical values of the scaling exponents $\beta$ for
different surface growth models on the different lattices
($\mathcal{H}$: Honeycomb, $\mathcal{S}$: Square, $\mathcal{T}$:
Triangular, $\mathcal{V}$:Voronoi). The numbers inside the
parenthesis are the error bars of the last digits.} \label{tab1}
\begin{center}
\begin{tabular}{lcccr}\hline\hline

& \multicolumn{4}{c}{Lattice type} \\
\cline{2-5}
Model  &$\mathcal{H}$  & $\mathcal{S}$ & $\mathcal{T}$ & $\mathcal{V}$ \\
\hline
RD&$0.500(1)$ &$0.500(1)$ &$0.500(1)$ &$0.500(1)$    \\
RSOS&$0.240(2)$&$0.240(5)$ &$0.240(5)$ &$0.240(3)$ \\
BD&$0.24(1)$ &$0.24(1)$ &$0.23(2)$ &$0.24(1)$

\\ \hline\hline

\end{tabular}
\end{center}
\end{table}

Our result confirms logarithmic scaling of roughness $W(t,L) \sim \ln (t)$ for RDR model where is in agreement with the predictions \cite{Barabasi,EW}.

\begin{itemize}
\item[b. ]\textit{Roughness exponent}
\end{itemize}

To measure the roughness exponent $\alpha$ we have calculated the
height-height correlation function $C(r)$ with respect to the
distance separation $r$  for saturated time.  Using numerical calculations for small
values of $r$, the roughness exponent can be read where Table
\ref{tab2} reports all the measured values of the roughness exponent
for BD and RSOS lattice growth models on different lattice types.
Results from numerical computation of $\alpha$ for growth processes
on the square lattice is in good agreement with the previously
reported values \cite{Barabasi,saberi}.
\begin{table}[htp]
\caption{ Numerical values of the scaling exponents $\alpha$ for BD and RSOS models on different lattice types.}
\label{tab2}
\begin{center}
\begin{tabular}{lcccr}\hline\hline

& \multicolumn{4}{c}{Lattice type} \\
\cline{2-5}

Model  &$\mathcal{H}$  & $\mathcal{S}$ & $\mathcal{T}$ & $\mathcal{V}$ \\
\hline
RSOS&$0.400(2)$& $0.400(2)$ &$0.400(3)$ &$0.400(5)$\\
BD &$0.30(4)$ &$0.29(4)$&$0.29(4)$ &$ 0.30(4)$
\\ \hline\hline

\end{tabular}
\end{center}
\end{table}

For RDR model deposited on different lattice types, we observed $C(r)\sim \ln (r)$ \cite{Barabasi,EW}.

\begin{itemize}
\item[c. ]\textit{Geometrical scaling exponents}
\end{itemize}

\begin{table}[htp]
\caption{ Numerical values of different geometrical exponents $x_l$,
$D_f$ and $\tau-1$ at $1\sigma$ confidence interval for the discrete surface growth models on
different lattices.} \label{tab3}
\begin{center}
\begin{tabular}{l|ccccr}\hline\hline

& &\multicolumn{4}{c}{Lattice type} \\
\cline{3-6}

   & &$\mathcal{H}$  & $\mathcal{S}$ & $\mathcal{T}$ & $\mathcal{V}$ \\
\hline
\multirow{4}{*}{$2x_l$}
&RDR&$1.0(1)$ &$1.0(1)$ &$1.0(1)$ &$1.0(1)$    \\
&RSOS&$1.0(1)$ &$1.0(1)$ &$1.0(1)$ &$1.0(1)$    \\
&RD&$0.53(5)$ &$0.47(5)$ &$0.49(4)$ &$0.52(5)$    \\
&BD&$1.0(2)$ &$1.0(2)$ &$1.0(2)$ &$1.0(2)$    \\
\hline
\multirow{4}{*}{$D_f$}
&RDR &$1.50(1)$ & $1.50(1)$ &$1.49(1)$ &$1.49(2)$   \\
&RSOS &$1.30(1)$ & $1.30(1)$ &$1.30(1)$ &$1.30(1)$  \\
&RD &$1.73(1)$ & $1.73(1)$ &$1.75(1)$ &$1.73(1)$    \\
&BD &$1.36(2)$ & $1.36(2)$ &$1.36(2)$ &$1.36(2)$    \\
\hline
\multirow{4}{*}{$\tau-1$}
&RDR&$1.33(1)$ &$1.34(1)$ &$1.34(1)$ &$1.33(1)$ \\
&RSOS&$1.22(1)$ &$1.23(1)$ &$1.22(1)$ &$1.22(1)$ \\
&RD&$1.18(1)$ &$1.19(1)$ &$1.19(1)$ &$1.19(1)$  \\
&BD&$1.32(2)$ &$1.31(2)$ &$1.32(3)$ &$1.32(2)$  \\
\hline
\multirow{4}{*}{$n$}
&RDR&$0.49(2)$ &$0.50(2)$ &$0.50(2)$ &$0.51(2)$    \\
&RSOS&$0.69(2)$ &$0.69(2)$ &$0.68(2)$ &$0.69(2)$    \\
&RD&$0.30(2)$ &$0.29(2)$ &$0.30(2)$ &$0.31(2)$    \\
&BD&$0.66(2)$ &$0.67(3)$ &$0.65(2)$ &$0.66(2)$    \\
\hline
\multirow{4}{*}{$m$}
&RDR&$0.02(3)$ &$0.02(2)$ &$0.02(2)$ &$0.02(2)$    \\
&RSOS&$0.03(3)$ &$0.03(3)$ &$0.03(3)$ &$0.02(3)$    \\
&RD&$0.03(3)$ &$0.02(3)$ &$0.02(2)$ &$0.03(3)$    \\
&BD&$0.04(5)$ &$0.04(4)$ &$0.03(4)$ &$0.03(3)$    \\
 \hline\hline

\end{tabular}
\end{center}
\end{table}

The most important exponent in fractal contour lines is the loop correlation function exponent $x_l$. For a given loop ensemble
 we followed the algorithm reported in the Ref. \cite{kondev1} to find the correlation function $G_c(r)$. 
The first part of the  Table .\ref{tab3} shows that the measured
values for the exponent $x_l$ for discrete growth processes i.e. RDR
model with the Gaussian statistics are in agreement with the
predicted value $2x_l=1$, within the statistical error. According to
our results, it is quite interesting that the relation $2x_1=1$ is
 valid also for non-Gaussian interfaces such as RSOS model. We
observed $2x_l=\frac{1}{2}$ for the RD model which is in agreement
with the predicted value for the two point correlation exponent of
the percolation model \cite{lolo}. The difference between RD model
and the other growth processes comes from the correlation between
lattice sites in the deposition process. There are no any
correlations between nearest neighbor sites in the RD model. The
same calculations are done for the BD model on different lattice
types. The large error bars which lead to the less  agreement in the
equality $2x_1 = 1$   come from finite size effects. One can compute
the perimeter, $s$, and gyration, $R$, radius of contour loops to calculate
fractal dimension $D_f$. Here $R$ is defined by $R^2=\frac{1}{N}\sum_{i=1}^{N}\left [(x_i-x_c)^2+(y_i-y_c)^2\right]$, with $x_c=\frac{1}{N}\sum_{i=1}^Nx_i$ and  $y_c=\frac{1}{N}\sum_{i=1}^Ny_i$. In the second part of the Table \ref{tab3}
we have shown the measured values of the fractal dimension of
contour loops for all surface growth processes on different lattice
types. The next remark concerns the probability distribution of
contour length $\tilde{P}(s)$ with the scaling exponent $\tau$. We
measured $\tau$ for different models, results are shown in the third
part of Table .\ref{tab3}. A suitable value of the parameter $\tau$
obtained form the best fit to $\ln(\tilde{P}(s))$ versus $\ln(s)$
(We measured $\tau -1$ instead of $\tau$). Finally we also calculated
$G_s(r)$ and the exponents $m$ and $n$; the results are listed in Table .\ref{tab3}.

 \begin{itemize}
\item[d. ]\textit{Scaling relations}
\end{itemize}

Finally the relation between measured scaling exponents $\alpha$,
$x_l$, $D_f$, $\tau$, $m$ and $n$ given by Eq. \ref{hyperscaling} and Eq. \ref{hyperscaling2} can be
examined. We have depicted the results for all growth models on
different lattice types in the Table .\ref{tab4}. As can be seen
from the Tables. \ref{tab1}, \ref{tab2} and \ref{tab3} the error
bars for the BD results are very large. It seems that to get
conclusive results one needs to simulate the BD model on very large
lattice sizes \cite{LargBD} which is very time consuming. However,
due to the finite size effect, we concluded that our results are
roughly consistent with the predictions.

\begin{table*}[htp]
\caption{ Verification of three basic hyper-scaling relations for
discrete surface growth process  on
different lattice types.} \label{tab4}
\begin{center}
\begin{tabular}{l|c||cc||cc||cr}
\hline
\hline
 & &  $D_f(\tau-1)$&$2-\alpha$&  $D_f(\tau-3)$& $2-2x_l$&$D_f(1-m)$&$2-n$ \\
\hline

\hline
\multirow{4}{*}{RD}
&$\mathcal{H}$&$2.04(3)$&$2$ &$1.42(3)$ &$1.47(5)$&$1.68(4)$&$1.70(2)$  \\
&$\mathcal{S}$&$2.06(5)$&$2$ &$1.40(5)$ &$1.53(5)$ &$1.69(4)$ &$1.71(2)$\\
&$\mathcal{T}$&$2.08(6)$&$2$ &$1.42(6)$ &$1.51(4)$ &$1.71(3)$ & $1.70(2)$ \\
&$\mathcal{V}$&$2.06(5)$&$2$ &$1.40(5)$ &$1.48(5)$ & $1.68(4)$& $1.69(2)$
\\ \hline

\multirow{4}{*}{BD}
&$\mathcal{H}$&$1.79(6)$&$1.70(4)$ &$0.93(4)$&$1.0(2)$ &$1.30(6)$ & $1.34(2)$ \\
&$\mathcal{S}$&$1.78(8)$&$1.71(4)$ &$0.94(6)$ &$1.0(2)$ & $1.30(6)$& $1.33(3)$\\
&$\mathcal{T}$&$1.79(6)$&$1.71(4)$ &$0.93(4)$ &$1.0(2)$ &$1.32(6)$ & $1.35(2)$\\
&$\mathcal{V}$&$1.79(6)$&$1.70(4)$ &$0.93(4)$ &$1.0(2)$ & $1.32(5)$ & $1.34(2)$

\\  \hline
\multirow{4}{*}{RDR}

&$\mathcal{H}$&$2.00(3)$ &$2$ &$1.01(3)$ &$1.0(1)$ &$1.47(4)$ & $1.51(2)$\\
&$\mathcal{S}$&$2.01(3)$&$2$&$0.99(3)$ &$1.0(1)$ &$1.47(3)$ & $1.50(2)$\\
&$\mathcal{T}$&$2.00(3)$&$2$&$0.98(3)$ &$1.0(1)$ & $1.46(3)$& $1.50(2)$\\
&$\mathcal{V}$&$1.98(4)$&$2$&$1.00(4)$ &$1.0(1)$ & $1.46(3)$& $1.49(2)$

\\ \hline
\multirow{4}{*}{RSOS}

&$\mathcal{H}$&$1.59(2)$&$1.60(1)$ &$1.01(2)$ &$1.0(1)$ &$1.26(4)$ & $1.31(2)$\\
&$\mathcal{S}$&$1.60(2)$&$1.60(1)$ &$1.00(2)$  &$1.0(1)$ &$1.26(4)$ & $1.31(2)$\\
&$\mathcal{T}$&$1.59(2)$&$1.60(1)$ &$1.01(2)$ &$1.0(1)$ & $1.26(4)$ & $1.32(2)$\\
&$\mathcal{V}$&$1.59(2)$&$1.60(1)$ &$1.01(2)$ &$1.0(1)$ & $1.27(4)$& $1.31(2)$
\\ \hline\hline
\end{tabular}
\end{center}
\end{table*}

\subsection{Conformal invariance (CI) test}
It is well-known, for a recent discussion see \cite{Rajabpour}, that
the action corresponding to Gaussian self-affine rough surfaces is
conformally invariant just for the surfaces with zero roughness
exponent. However, recently many authors claimed that despite the
non-conformal invariant height ensembles of these surfaces, their
contour lines might be conformally invariant
\cite{turbulence2,Niry,saberi}. In the following we examine this
guess by using the direct conformal invariance test used first in
\cite{langlands1} . Suppose a conformal mapping defined by
$w=g_t(z)$. For any domain $\mathcal{D}$ with boundary
$\mathcal{C}$, one can find a conformal map $g_t$ where maps
$\mathcal{D}$ to $\mathcal{D^{\prime}}$ and  $\mathcal{C}$ to
$\mathcal{C^{\prime}}$. For the critical statistical systems, the measure on
distributions on $\mathcal{C^{\prime}}$  is obtained by transport of
the measure on distributions on $\mathcal{C}$ using  conformal
mapping $g_t$ from $\mathcal{D}$ to $\mathcal{D^{\prime}}$ where
this measure is invariant at the critical point of the model.

Statistical systems such as, percolation and Potts models at the
critical point, exhibit a unique spanning cluster, where connect the
boundaries of domain from one side to the other (i.e. left to right
or up to down) along the spanning cluster \cite{stauffer}. In the
scaling limit when the size of the system goes to infinity, one can
define the crossing probability $\pi_h$ ($\pi_v$), the probability
of a system to percolate only in the horizontal (vertical) direction
\cite{langlands1}. In the conformal invariant statistical systems,
the measure $\pi_h$ ($\pi_v$) is unchanged under any conformal map
$g_t$ where it maps a given boundary $\mathcal{C}$ to
$\mathcal{C}^\prime$
\cite{langlands1,langlands2,langlands3,blanchard}.

The conformally invariant curves should be statistically equivalent
to Schramm Loewner evolution (SLE$_\kappa$). In this method the
conformal map $g_t(z)$ which maps the half-plane minus the trace
$\gamma_t$ into itself, obeys the Loewner's formula $\partial_{t}
g_{t} =2/{(g_t-\xi_t)} $ where the driving function $\xi_t$ is
related to the Brownian motion with the diffusivity $\kappa$. The
fractal dimension of such scaling curves
 is given by the relation $d_f=1+\kappa/8$. In \cite{turbulence2,Niry,saberi}, using a discretized Loewner equation and iterative conformal slit
 map they extracted the drift $\xi_t$ for iso-height lines in the different rough surfaces. They found that
 an ensemble of the driving function $\xi_t$ for these models looks like converging to a Gaussian process with
 zero mean $\langle \xi_t \rangle \approx 0$ and the variance $\langle \xi_t^2 \rangle \approx \kappa t$ with
 $\kappa $ compatible with the fractal dimension of the curves. It shows for example that the model with logarithmic height correlation
  can be related to SLE$_4$.
  However, it seems that the
 calculations based on Loewner evolution (using Schramm Loewner equation to find $\kappa$) are not accurate enough. Especially it is not easy
 to find the exact Gaussian distribution for the increments of the driving function and also it is difficult to show stationarity of increments
 of the drift  to conclude that the drift is really a Brownian
 motion. For example even for RDR (which is a Gaussian free field) we know that usually the drift is
 not just a Brownian motion, it is a complicated function related to
 the Brownian motion\cite{GFF}.

An alternative way is to apply directly the conformal invariant test
 to a two dimensional slice of random rough surface. For a given
rough surface sample $h(x,y)$, a horizontal cut is made at a certain
level
 $h_{c}=\langle h\rangle +\delta \sqrt{\langle [h-\langle h\rangle]^2 \rangle}$,
  where the symbol $\langle ...\rangle$ is averaging over ensemble of rough surfaces.
  A set of nearest-neighbor connected sites of positive (negative) height form a cluster.
  The cluster is spanning one if spans two opposite sides of a given domain, such that infinite connectivity first occurs,
  and one can measure the crossing probabilities $\pi_h(\delta)$ and $\pi_v(\delta)$ for a given domain geometry.

 In this study we considered square boundary as $\mathcal{C}$ and rhombus domains with different angle  as $\mathcal{C}^\prime$
. There is a conformal map that takes interior of our choice of
$\mathcal{C}$ to the interior of $\mathcal{C}^\prime$ and
 takes vertices to vertices and sides to sides. The basic method to perform this test composed of three steps: (I)
 draw the boundary $\mathcal{C}$ ( $\mathcal{C^\prime}$)  defining cluster ensemble on the lattice.
 (II) assign a state $+1$ ($-1$) to each site of the lattice  located inside $\mathcal{C}$ ($\mathcal{C^\prime}$) when $h(x,y)>h_c$ ($h(x,y)\leq h_c$). (III)
 find cluster of nearest-neighbor connected sites with the same values (for example $+1$) by using Hoshenn-Kopelman (HK) algorithm \cite{HK}.
 These three steps are repeated for all of the configurations of level set ensemble for various $\delta$'s to find the largest cluster
 where spans left and right, to measure $\pi_h$ (up and down to measure $\pi_v$).
The expected value of $\pi_h$ ($\pi_v$), is then the ratio of the number of configurations spanning two
 opposite horizontal or vertical sides of the boundary $\mathcal{C}$ ( $\mathcal{C}^\prime$) to the sample size.

We examined the above test for two dimensional slice of RD, RDR,
RSOS and also synthetic (2+1)Dimensional self-affine surfaces which known as
fractional Brownian motion (2D FBM) \cite{smvaez} for different
$\delta$'s and we measured crossing probability $\pi_h(\delta)$
($\pi_v(\delta)$) for
 different  boundaries  $\mathcal{C}$ (square with side length $L$) and $\mathcal{C}^\prime$ (rhombus with angle $\theta$  and side $L$).
  In order to have an ensemble of surface height profiles, we have obtained $2\times10^{5}$ samples from the grown rough surfaces with side $L=512$
  for each model. We measured the probabilities $\pi_h(\delta)$ and $\pi_v(\delta)$  that at each level height $\delta$, an infinite
  island spans two opposite boundaries of the domains.

\begin{figure}
\begin{center}
\includegraphics[width=0.9\linewidth]{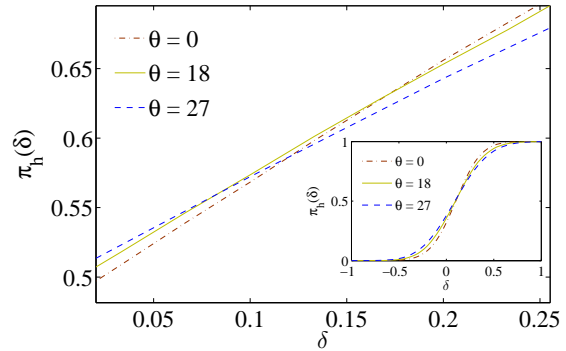}\\
\caption{\label{fig.3} (Color online)
 The spanning probability $\pi_h(\delta)$ for two dimensional slice of RSOS model on rhombus domain with different angles $\theta$. Inset: The same quantity for RDR.}
\end{center}
\end{figure}

As shown in Fig \ref{fig.3} the measured  crossing probabilities for
RDR model on rhombus domains with different angles $\theta$,
 cross each other at $\delta = \delta_c$, implying that the crossing probabilities remain constant by conformal transformation.
 This means that the two dimensional slices of RDR model at the critical level $\delta_c$ is possibly conformally invariant
 and the crossing probabilities do not change by conformal transformations. We also applied this test to 2D FBM
 with Hurst exponent $0\leq H \leq 1$ and we observed conformal invariance only for $H=0$.
 We observed that the 2D FBM with $H>0$ is conformal non-invariant and the spanning probability $\pi_h(\delta)$
 do not show any fixed point. In addition we presented in Fig.  \ref{fig.3} the  curves $\pi_h(\delta)$ obtained for RSOS model on different boundaries.
 It shows that the crossing probability for two dimensional slice of RSOS model for different $\delta$'s,
 change by conformal transformation of the boundary. Increasing the number of ensembles and the size of the system did not make significant change in this pattern.
   We concluded that  this model is also conformally non-invariant.

Our results show that the contour ensemble of RD model on the triangular
lattice at the mean height ($\delta=0$) is conformally invariant.
Two dimensional slice of RD model at the mean level for other
lattice types is conformal non-invariant. This is obvious if we
think to the clusters of RD as a simple percolation clusters. 
Our numerical results are summarized in Tabel. (\ref{tab6}).

\begin{table}[htp]
\caption{ CI test results of two dimensional loop ensembles of
random rough surfaces.} \label{tab6}
\begin{center}
\begin{tabular}{lcccr}\hline\hline

& \multicolumn{4}{c}{Lattice type} \\
\cline{2-5}

 model   & $\mathcal{S}$ & $\mathcal{T}$  \\
\hline
RD &No &Yes     \\
RDR  &Yes & Yes   \\
RSOS &No &No\\
FBM$(H\neq 0)$&No&No
\\ \hline\hline

\end{tabular}
\end{center}
\end{table}

\section{Conclusion}
In this paper we showed that many properties of rough surfaces such
as the roughness exponent and dynamical exponent  are independent of
the underlying lattices, see Table. \ref{tab1} and \ref{tab2}. We
also showed that the fractal dimension and correlation exponent and
many other exponents of contour lines of the surfaces are also
independent of the underlying lattices and the distribution of the heights. 
All the calculations were
carried out on four different regular and irregular lattices such as
the square, triangular, honeycomb and Voronoi lattices, see Table.
\ref{tab3}. Although the calculations for BD were not conclusive we
were able to show that the extracted critical exponents of the
contour lines of RDR, RSOS and RD follow the hyperscaling relations independent of
having a Gaussian or non-Gaussian surface,
see Table. \ref{tab4}. Finally using direct conformal maps we showed
that the contour lines of self-affine rough surfaces (RSOS, BD and
fractional Gaussian rough surfaces) with non-zero roughness
exponents are not conformally invariant. To be specific just the
contour lines of RDR on different lattices and RD on the triangular
lattices are conformally invariant, see Table. \ref{tab6}.
\newline

\textbf{Acknowledgment:}
We thank M. Ghasemi Nezhadhaghighi for many discussions and helps. We also thank  A Saberi and S Rouhani
for reading the manuscript. The work of S. Hosseinabadi was supported by Islamic Azad University,
East Tehran Branch. MAR thanks FAPESP for finantial support.

\end{document}